\documentclass[11pt,a4paper]{article}
\usepackage{graphicx}
\usepackage{latexsym}
\usepackage{amsmath,amssymb,amsthm}
\usepackage{dsfont}
\usepackage{amsfonts}
\usepackage{algorithm}
\usepackage{algpseudocode}
\usepackage{placeins}
\usepackage[misc]{ifsym}
\usepackage{longtable}
\usepackage{hyperref}
\usepackage{listings}
\usepackage{float}
\date{}

\begin{document}
	\providecommand{\abs}[1]{\lvert#1\rvert}{\tiny {\tiny {\large }}}
	\providecommand{\norm}[1]{\lVert#1\rVert}
	\newcommand\sIf[2]{ \If{#1}#2\EndIf}
	\renewcommand\thealgorithm{\thesection.\arabic{algorithm}} 
	\newcommand{\bea}{\begin{eqnarray}}
		\newcommand{\eea}{\end{eqnarray}}
	\newcommand{\nn}{\nonumber}
	\newcommand{\bee}{\begin{eqnarray*}}
		\newcommand{\eee}{\end{eqnarray*}}
	\newcommand{\lb}{\label}
	\newcommand{\nii}{\noindent}
	\newcommand{\ii}{\indent}
	\newcommand{\etal}{\textit{\etal}. }
	\newcommand{\ie}{\textit{i}.\textit{e}., }
	\newcommand{\eg}{\textit{e}.\textit{g}. }
	\newtheorem{thm}{Theorem}[section]
	\newtheorem{cor}{Corrolary}[section]
	\newtheorem{lem}{Lemma}[section]
	\newtheorem{assumption}{Assumption}[section]
	\newtheorem{prop}{Proposition}[section]
	\newtheorem{rem}{Remark}[section]
	\newtheorem{defn}{Definition}[section]
	\numberwithin{equation}{section}

\title{The impact of the  additional features on the performance of regression analysis: a case study on regression analysis of	music signal}

\author{V. N. Aditya Datta Chivukula \thanks{Department of Computer Science, International Institute of Information Technology 	Bhubaneswar, India 751003;
		E-mail: B518017@iiit-bh.ac.in }   \and Rupaj Kumar Nayak \noindent
	\thanks{Corresponding author. Department of Mathematics,International Institute of Information Technology, Bhubaneswar, India; E-mail: rupaj@iiit-bh.ac.in}}

\maketitle \noindent

\begin{abstract}
Machine learning techniques nowadays play a vital role in many burning issues of real-world problems when it involves  data. In addition, when the task is complex, people are in dilemma in choosing deep learning techniques or going without them. This paper is about whether we should always rely on deep learning techniques or it is really possible to overcome the performance of deep learning algorithms by simple statistical machine learning algorithms by understanding the application and processing the data so that it can help in increasing the performance of the algorithm by a notable amount. The paper mentions the importance of data preprocessing than that of the selection of the algorithm. It discusses the functions involving trigonometric, logarithmic, and exponential terms and also talks about functions that are purely trigonometric. Finally, we discuss regression analysis on music signals to justify our claim.
\end{abstract}

\section{Introduction}
Regression analysis gained its importance when several statisticians found out its applications in the real-world such as predicting the price of land in a certain city, estimating the complex polynomials through working on the dataset provided, estimating whether a given medicine will work on a large amount of people etc.. The research by Yao et al. \cite{yao2005functional}  gained its importance during the past decade with its description of solving various statistical models. 

Although trigonometric functions based estimation alone is known to suffer from bias problems at the boundaries due to the periodic nature of the fitted functions, Eubank, and Speckman \cite{eubank1990curve} presented a method of estimating an unknown regression curve by regression on a combination of low-order polynomial terms and trigonometric terms. 

A lot of  literature are available on the primary variations of regression of which few are listed in \cite{liaw2002classification,ostertagova2012modelling,zou2003correlation}. These algorithms have their own importance individually and are application-specific. Therefore, the practical realization of technical research applications needs their respective algorithms or approaches which has better efficacy and improves the accuracy of the applications with the least error possible. 

There are some areas where we need to understand the importance and need for a perfect combination of above mentioned requirements in a simple way to enhance the accuracy of results. Also, we need to understand the true efficiency of regression analysis in many other fields which are quite recent with respect to the growing demand of new applications in research.
 
Motivated on the approach by Eubank, and Speckman \cite{eubank1990curve} and need for showing the application of regression analysis on some complex tasks, we discuss  about trigonometric regression and polynomial regression on hypothesis involving logarithmic or exponential terms to establish the importance of adding features to the dataset for better results. Further, the research provides the contrast between the performance delivered by the above mentioned methods and simple neural networks. The main contribution  of the paper is that a proper data pre-processing  step can highly reduce the error and allows someone to solve problems with much more light-weight and making the method basic. Hence for establishing the claim, the complex tasks like music signal analysis is considered for experiment. 

\section{Regression analysis of the trigonometric function}
We discuss the regression analysis of the trigonometric function in this section. For this, we generate a trigonometric function randomly using a python code given in listing \ref{list1}. 
\lstset{language=Python}
\lstset{frame=lines}
\lstset{caption={Python Code for generating function with only trigonometric terms}\label{list1}}
\lstset{label={lst:code_direct}}
\lstset{basicstyle=\footnotesize}
\begin{lstlisting}
	feature = ['x','np.sin(x)','np.cos(x)','np.sin(x)*np.cos(x)']
	function = []
	for i in range(len(feature)):
	coef = str(np.random.choice(np.ara-nge(100)))
	term = coef + '*' + np.random.cho-ice(feature[1:])
	function.append(term)
	function = '+'.join(function)
	function = 'y=' + function
\end{lstlisting}

In the code, there is a feature list containing all features of our interest. There is a single `For' loop ranging from $0$ to the length of feature list. An individual is allowed to choose a range which is equal to the number of terms that are required in the end polynomial. For each iteration of the loop, we randomly select coefficient for each term and the term itself from the feature list. Then, we multiply the coefficient and store the resulting string in a list known as function. We continue the same until the loop is completed. Hence, we end up having a list of terms as strings. Finally, we join all the strings using `join' function which results in a random trigonometric polynomial in string datatype. Noted that range of loop is the number of terms one desires in the end function and the feature $x$ is not considered while generating the function as  this section is devoted towards a pure trigonometric function.

The  trigonometric function generated is:
\begin{align}\label{1}
y=95\sin x\cos x+37\sin x+90\sin x\cos x +45\sin x\cos x.
\end{align}

\begin{figure}[htp]
\begin{minipage}[b]{.5\textwidth}
		\centering
		\includegraphics[width=.75\textwidth]{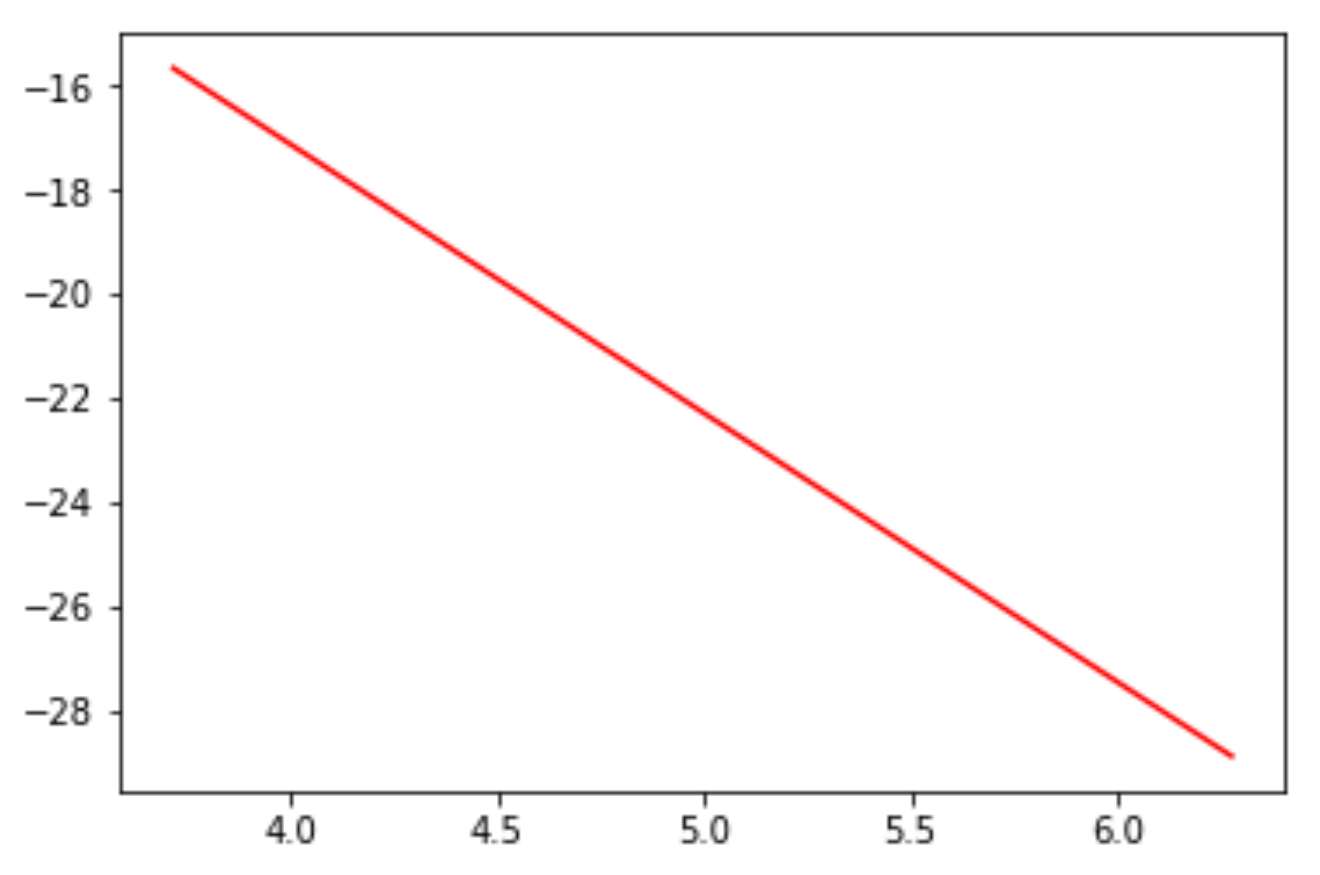}
		\caption{Predictions on y-axis with inputs\\ on x-axis for simple linear regression}\label{fig1}
	\end{minipage}%
	\begin{minipage}[b]{.5\textwidth}
		\centering
		\includegraphics[width=.75\textwidth]{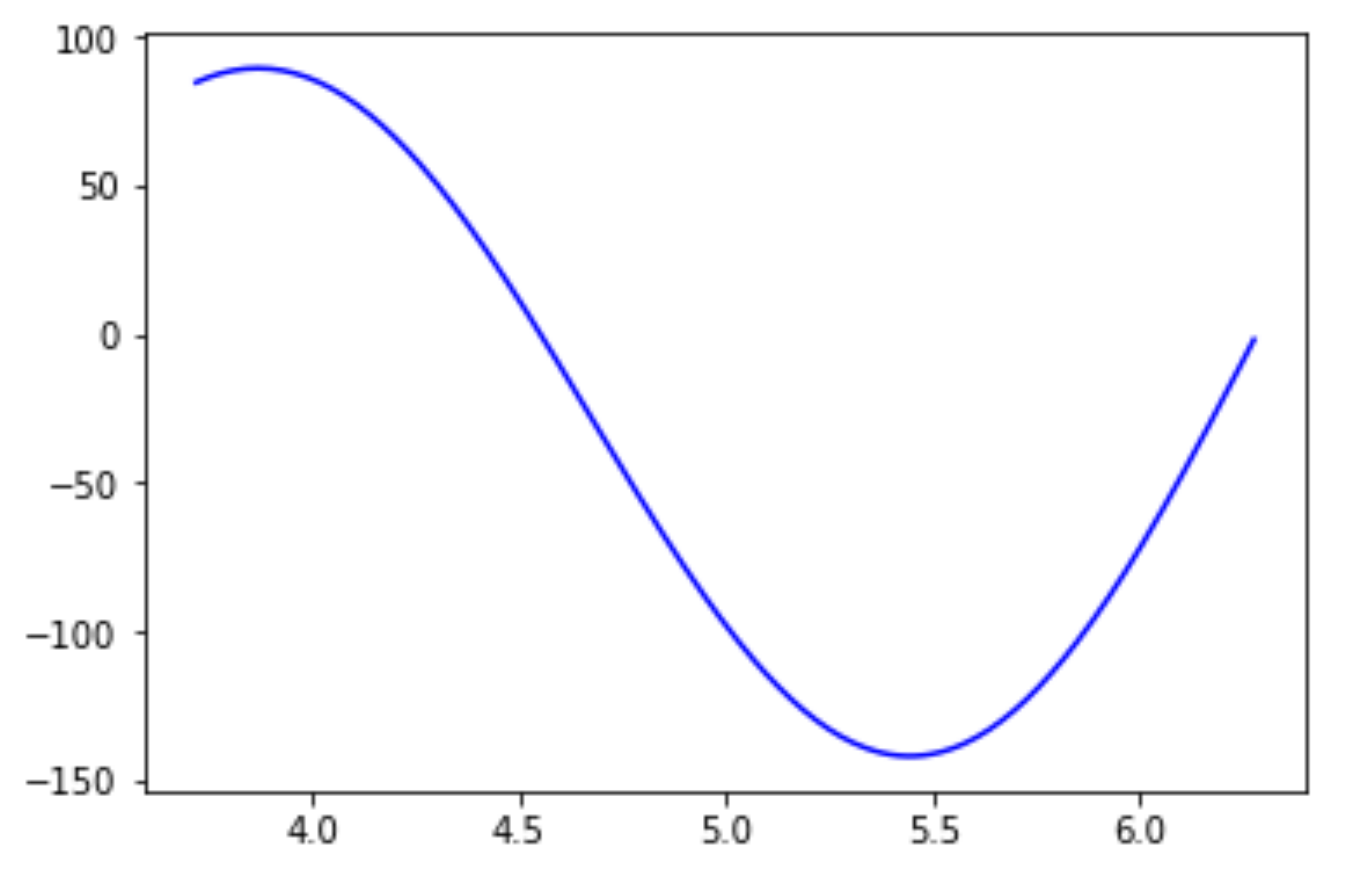}
		\caption{Plot depicting desired outputs for the inputs}\label{fig2}
	\end{minipage}
	\begin{minipage}[b]{.5\textwidth}
		\centering
		\includegraphics[width=.75\textwidth]{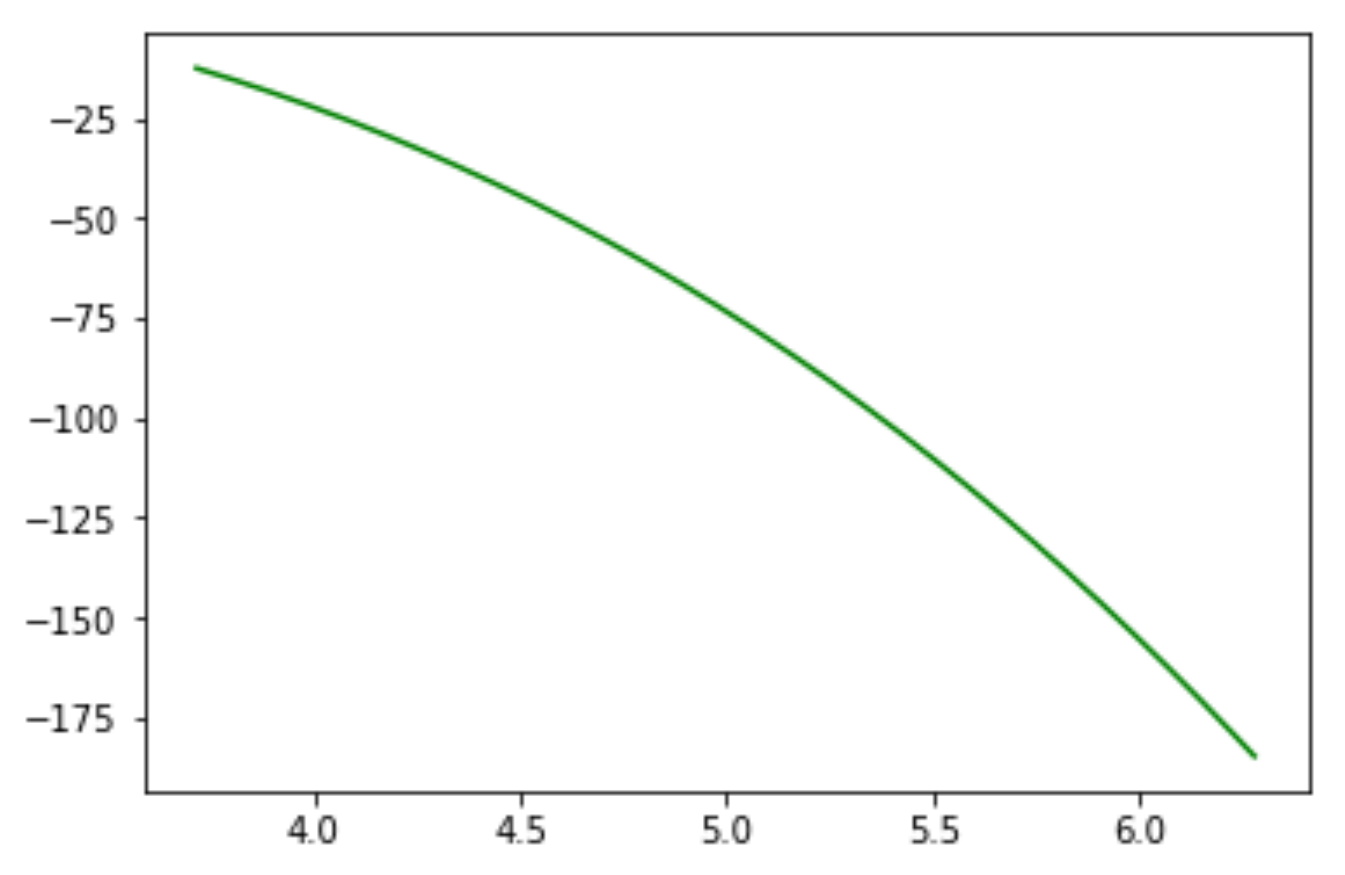}
		\caption{Predictions on y-axis and inputs\\ on x-axis for polynomial regression}\label{fig3}
	\end{minipage}
	\begin{minipage}[b]{.5\textwidth}
	\centering
       \includegraphics[width=.75\textwidth]{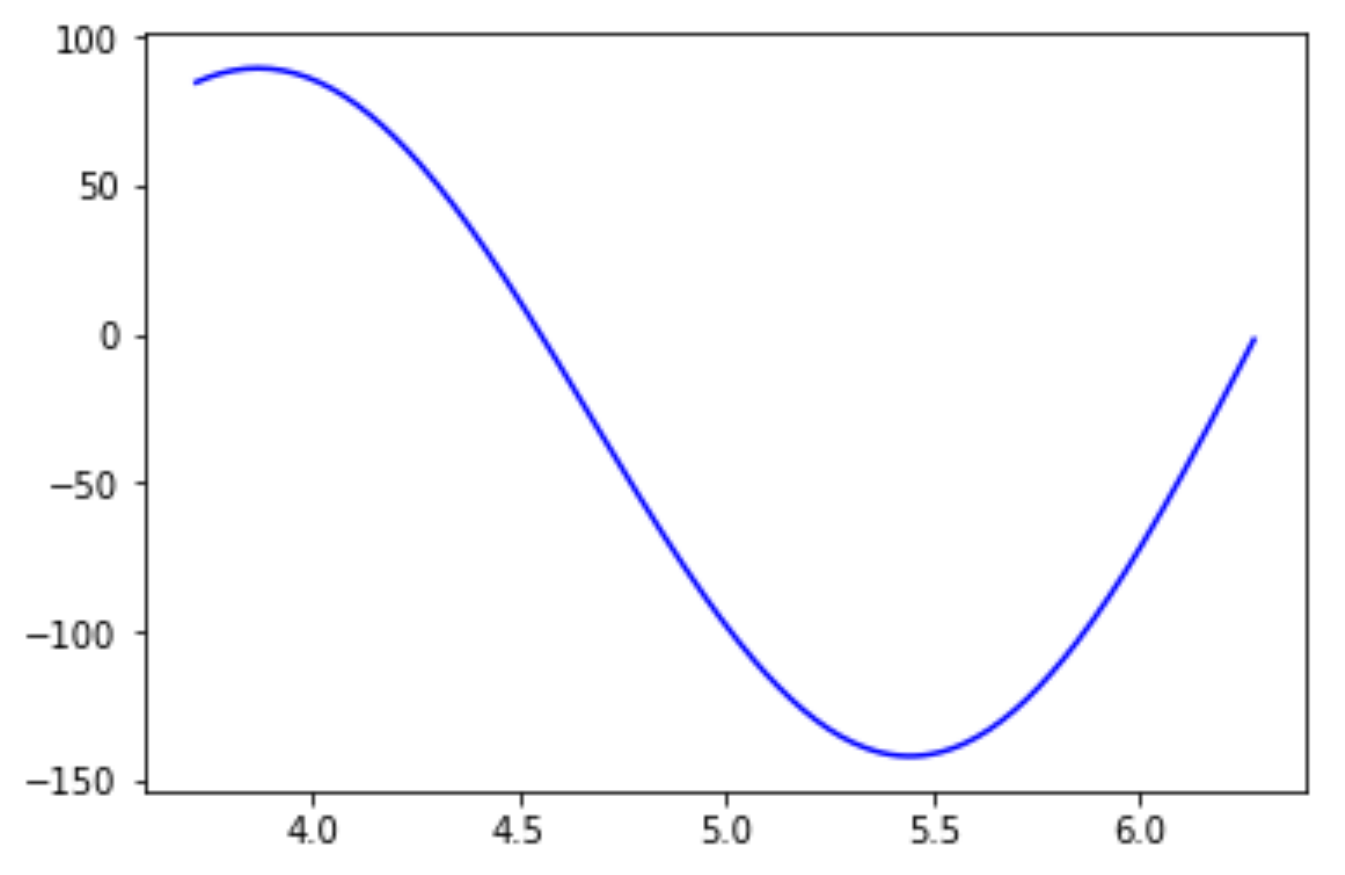}
       	\caption{Plot with inputs on x-axis and predictions by simple linear regression after adding trigonometric features to the dataset on y-axis}
		\label{fig4}
	\end{minipage}
\end{figure}

Equation (\ref{1}) is the function taken to explain the importance of trigonometric features in regression analysis. Noted that, there are no terms with $x$ raised to a certain power.  When we apply linear regression analysis on the dataset with input as $x$, with  $x\in [-\pi, \pi]$ in steps of $0.01$, the output $y$ is calculated  for thousand samples. The graph shown in fig.\ref{fig1}  depicts the performance of the linear regression on the test set, whereas  the desired performance is shown in fig.\ref{fig2}. Hence, we can decide that the linear regression performed poorly as expected. Now, if we use a polynomial regression and consider the hypothesis degree to be $2$ and train on the same training data and test it, we obtain performance as shown in fig.\ref{fig3}. It is expected that the polynomial regression cannot predict the trigonometric terms as there is no feature which is trigonometric in nature. 

Now, one can always think about using a simple neural network. However, that also would not work, as the training set is too low for the neural network to generalize the trigonometric hypothesis. Further, training the network excessively for a greater number of epochs would result in overfitting of data and also does not assure accuracy \cite{goodfellow2016deep}. We can also try with Long short-term memory (LSTM) \cite{hochreiter1997long} to overcome the issues. But, we should not forget the fact that LSTM networks require a high amount of data and moreover are computationally expensive as compared to the simple neural networks and regression analysis discussed above. 

Now, if we introduce the trigonometric terms in the hypothesis considered in the case of simple linear regression as redefined according to equation (\ref{1}) and train on the dataset with a new hypothesis and apply linear regression analysis then we can observe the performance as shown in fig.\ref{fig4}. Thus, by looking at fig.\ref{fig2} and fig.\ref{fig4}, one can understand the importance of trigonometric features in linear regression provided the dataset has a trigonometric relationship. The absolute error given in the Table \ref{tab1} shows the errors obtained with each regression approach discussed.

\begin{table}[H]
	\caption{Error table for pure trigonometric function by different algorithmic approaches.}
	\label{tab1}
		\begin{center}
			\begin{tabular}{|l|c|c|c|r|}
					\hline
					Algorithm & Absolute error \\
					\hline
					Proposed approach    & 6.610267888618182e-12  \\\hline
					Linear Regression    & 18573.351509906905 \\\hline
					Polynomial Regression & 15689.82990204867 \\
					\hline
				\end{tabular}
		\end{center}
\end{table}
Generally, the need of trigonometric regression analysis is felt in the fields of signal processing and wave analysis. Hence we want to study further on this approach. In section 3, we discuss polynomial trigonometric regression where we consider adding trigonometric features as a primary data preprocessing step whenever we encounter with regression analysis problems.
\section{Regression analysis of polynomial with trigonometric features}
In section  2 we have discussed function having only trigonometric terms without the mixture of linear or quadratic terms in $x$, where $x$ is the input value.  It may be noted that a function that contains a term like $x\cos x$ and so on, is difficult for simple neural networks and even the simple statistical regression algorithms like linear regression and polynomial regression to learn on minimal data. This attracts us to discuss further in this section.

Equation (\ref{2}) is generated using the code provided by listing \ref{list2}. To briefly explain the algorithm, in the first loop the degree of the polynomial is kept as range and all orders of input feature $x$ are included in the features list. Then, every term in the `terms' list is included in the features list. Now, when the `features' list is ready, a `function' is defined, in which, an empty list `T' is considered and the number of terms in the generated polynomial is decided at random by keeping a maximum upper-limit. Now, a loop is considered keeping number of terms as range and for each iteration, a term is appended to list `T' by generating the term with a randomly selected number of features. Finally, polynomial is created by joining the terms stored in list `T'.
\newpage
\lstset{language=Python}
\lstset{frame=lines}
\lstset{caption={Python code to generate a random mixed polynomial}\label{list2}}
\lstset{label={lst:code_direct}}
\lstset{basicstyle=\footnotesize}
\begin{lstlisting}
	x = np.pi # buffer value
	functions = []
	terms = ['np.cos(x)','np.sin(x)','np.tan(x)','np.log(x)','np.exp(x)']
	features = []
	for i in range(2):
	features.append("x**"+str(i+1))
	for i in terms:
	features.append(i)
	
	# generating function
	def function():
	T = []
	number_terms = np.random.cho-ice(np.arange(10))+1
	for i in range(number_terms):
	num_features = np.random.cho-ice(len(features))+1
	l = []
	for j in range(num_features):
	l.append(features[np.random.cho-ice(np.arange(len(features)))])
	t = '*'.join(l)
	T.append(t)
	func = '+'.join(T)
	func = 'y='+func
	return func
\end{lstlisting}

 The function  thus generated is: 
\begin{align}\label{2}
	y = e^x \cos x \tan^2 x &+x^3 \sin x + x^3 \tan x \sin x \log x\\ &+ x^3+ x^3 \cos x \tan x e^x \log x+ x^4 e^x \tan x. \notag
\end{align}
We observed that terms in the  equation \ref{2} containing product of algebraic and trigonometric functions (without simplification of trigonometric terms).  Here we study the inclusion of the additional features  including trigonometric, logarithmic and exponential features in $x$ and also all possible permutations of them. Once the individual estimates the degree of polynomial, the learning hypothesis performs the same way as we do in case of normal polynomial regression.

\begin{figure}[htp]
		\begin{minipage}[b]{.5\textwidth}
		\centering
		\includegraphics[width=.8\textwidth]{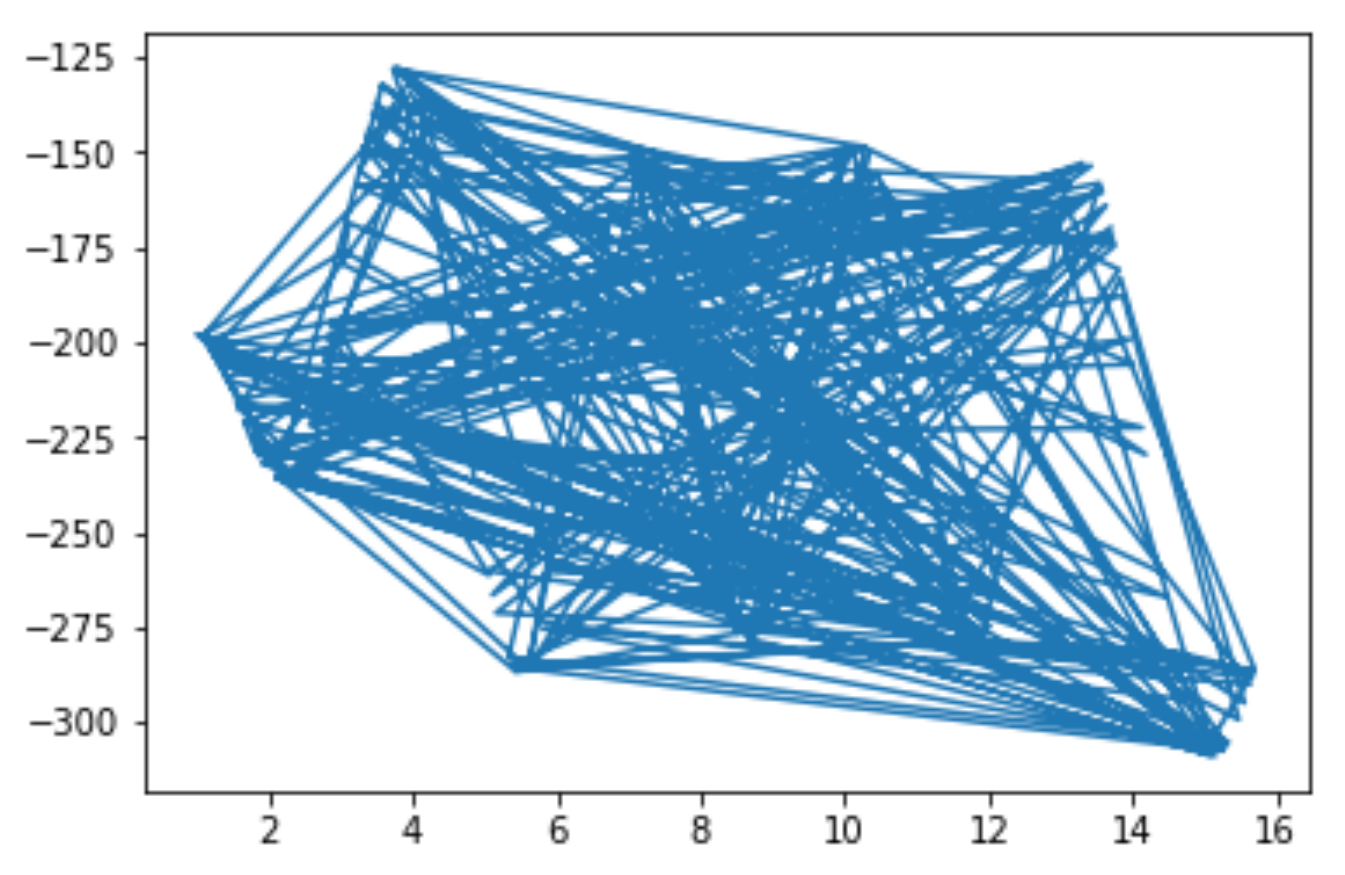}
		\caption{Plot depicting predictions on y-axis\\ and input value on x-axis by support vector\\ regression}\label{fig5}
	\end{minipage}%
	\begin{minipage}[b]{.5\textwidth}
		\centering
		\includegraphics[width=.79\textwidth]{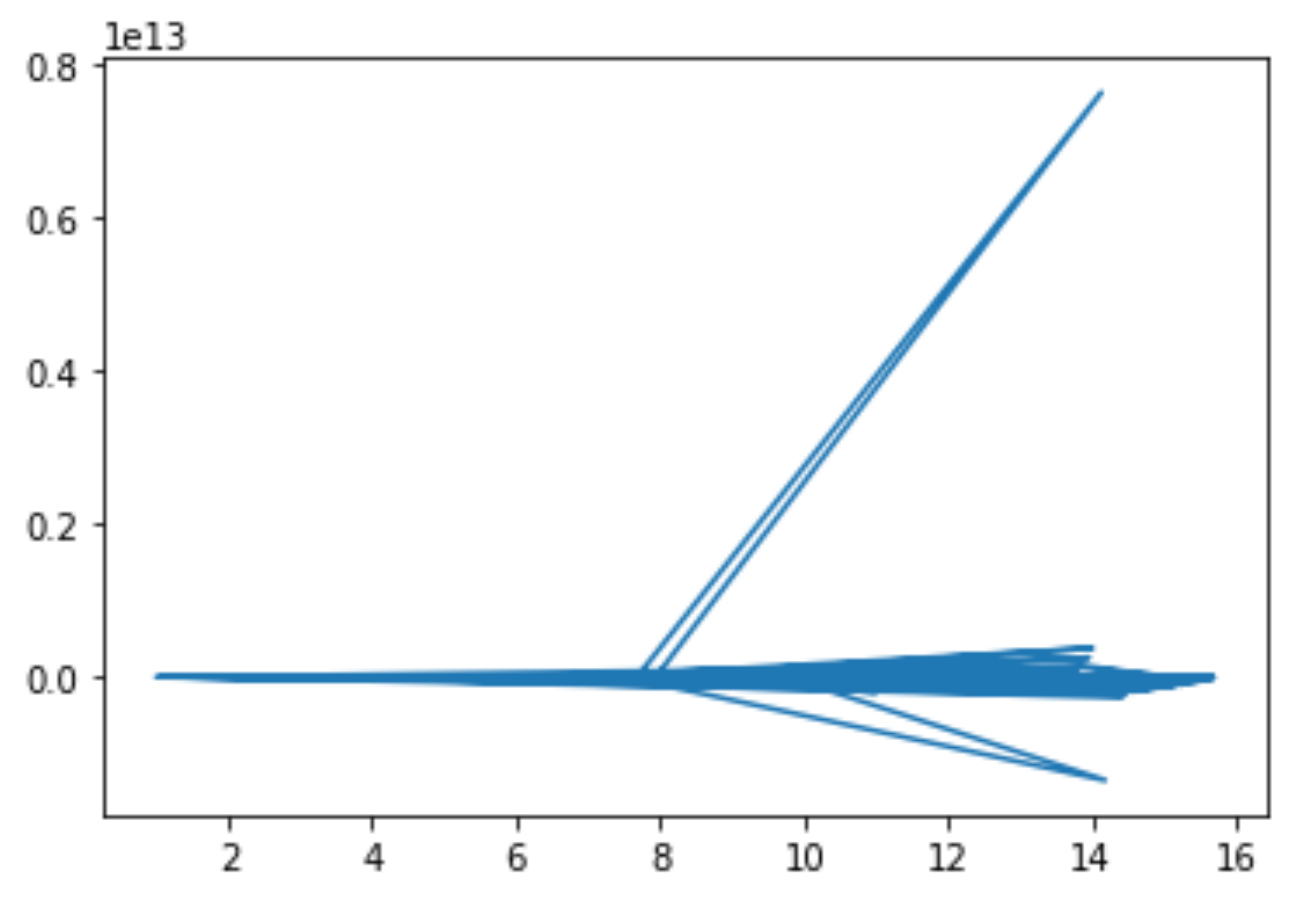}
		\caption{Plot depicting expected outputs on y-axis for inputs on x-axis}\label{fig6}
	\end{minipage}
	\begin{minipage}[b]{.5\textwidth}
		\centering
		\includegraphics[width=.85\textwidth]{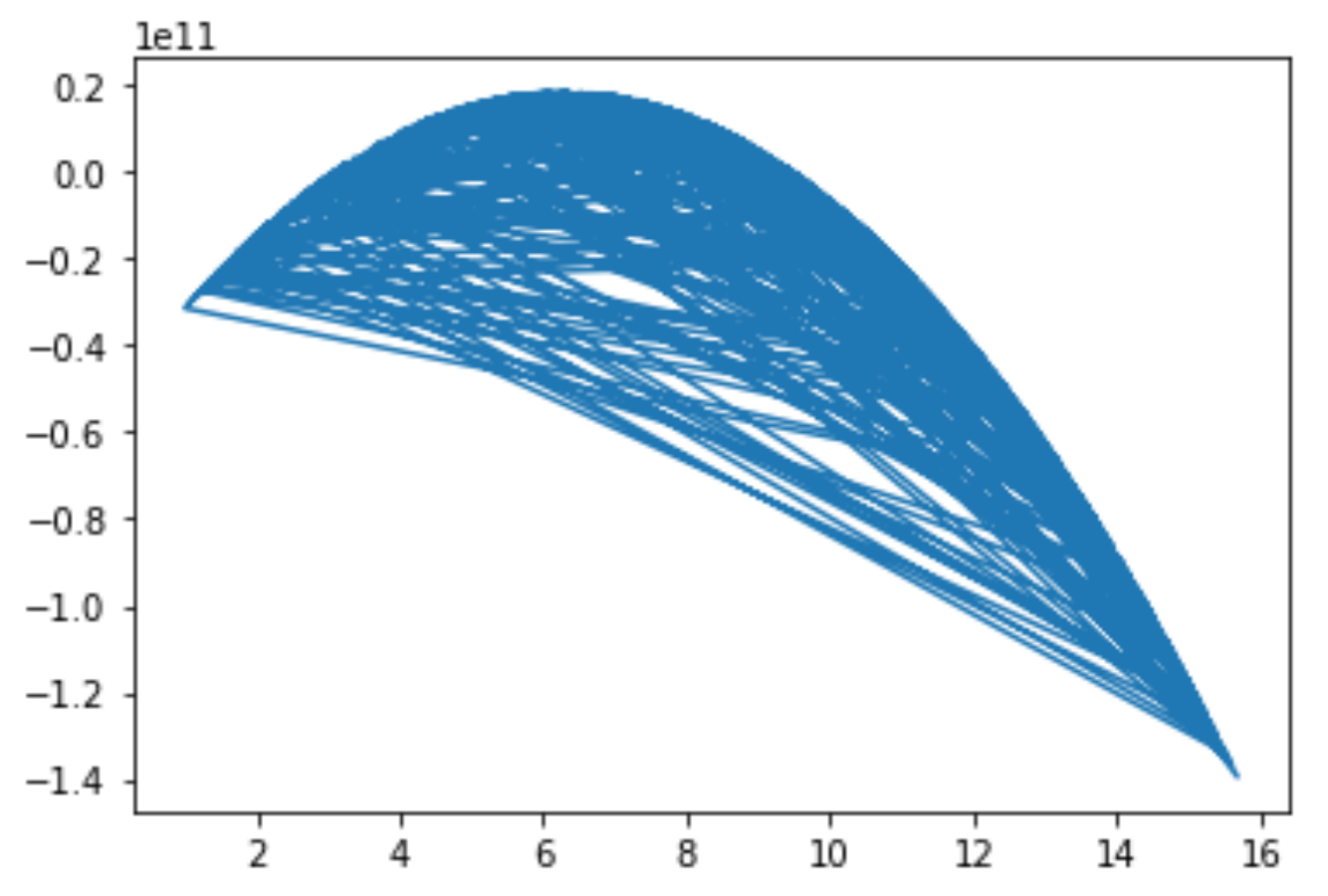}
		\caption{ Plot depicting predictions on y-axis\\ for inputs on x-axis by polynomial regression}\label{fig7}
	\end{minipage}
\begin{minipage}[b]{.5\textwidth}
	\centering
	\includegraphics[width=.79\textwidth]{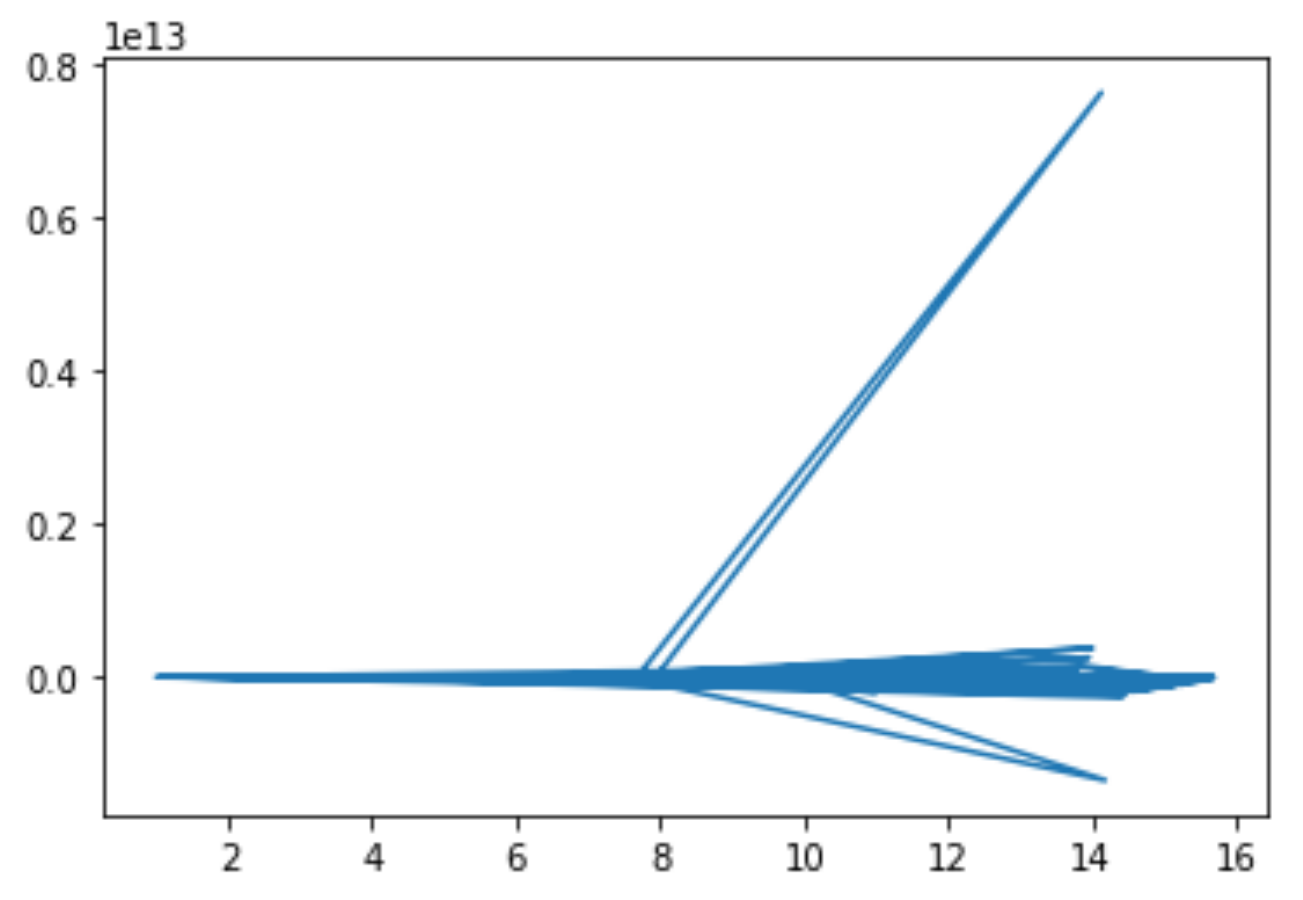}
	\caption{ Plot depicting predictions on y-axis for inputs on x-axis by linear regression after addition of features }\label{fig8}
\end{minipage}
\end{figure}

The figures \ref{fig5}, \ref{fig6}, \ref{fig7} and \ref{fig8} depicts different predictions analysis by different algorithms. 

If we carefully observe fig.\ref{fig5} which depicts the predictions by support vector regression trained on dataset with inputs ranging from $-\pi$ to $\pi$ and outputs calculated according to equation \ref{2}, we see that the expected plot as in fig.\ref{fig6} is completely different from what has been predicted which leads to high absolute error on test set. When we apply polynomial regression analysis keeping the degree as 2, then also we can see that the plot by polynomial regression as depicted in fig.\ref{fig7} is mostly off in predicting the desired outputs as shown in fig.\ref{fig6}.

Hence, if we are able to actually consider the list of additional features which are all possible permutations of $x$ with trigonometric, logarithmic and exponential functions acting upon it and then apply linear regression analysis, we observe the desired plot as in fig.\ref{fig8} which is almost similar to actual relationship showcased in equation \ref{2}. The errors in Table \ref{tab2} justifies our claim.
\begin{table}[!h]
	\caption{Error table for polynomial with complex terms by different algorithmic approaches}
	\label{tab2}
	\begin{center}
				\begin{tabular}{|l|c|c|c|r|}
					\hline
					Algorithm & Absolute error \\
					\hline
					Proposed approach    & 27.97901221743491  \\\hline
					Support Vector Regression    & 14177902477532.947 \\\hline
					Polynomial Regression & 15.715957$e$+12 \\
					\hline
					\end{tabular}
		\end{center}
\end{table}
Comparing fig.\ref{fig6} and fig.\ref{fig8}, one can conclude that the simple addition of all combination of functional features can affect the performance of an algorithm by a great extent. Table \ref{tab2} depicts the errors obtained by discussed algorithms.
If one thinks that the number of permutations is increasing with the degree of the hypothesis then by applying the dimensionality reduction techniques, the computational time can be decreased. This approach is only successful when the input is related to output with assumed combinations of features. We can also analyze data in preprocessing stage to identify more complex functions as features in $x$ depending upon the dataset.

\section{Music signal analysis}
Music signal is one of the complicated signals on which an efficient machine learning algorithm also suffers in learning  the parameters such as amplitude, frequency and phase as the superposition of several sinusoidal waves change after very short span of time over the complete time interval. Assuming that there are only a constant number of waves superposed over each short span of time frame and consider a superposition as shown in the following equation:
\begin{equation}\label{3}
	y = \sum_{i=0}^{20} a_i\sin(2\pi(f_i(x) + phase_i))
\end{equation}
where,
\begin{align*}
	a_i &= \text{amplitude parameter of} ~~i^{th}~ \text{wave}\\
	f_i &= \text{frequency parameter of}~ i^{th}~ \text{wave}\\
	phase_i &= \text{phase parameter of }~i^{th}~ \text{wave}\\
\end{align*}

Then we can optimize the parameters using many optimization algorithms. However,  we have taken the gradient descent algorithm to optimize which is simple to  apply. Here, we considered a random background music track \cite{gopisundar2020charlie} for explanatory purpose and considered first 800,000 samples of the audio amplitudes from left channel, then, we have further divided the entire training set into $800$ segments with each containing $1000$ samples. These $1000$ samples are trained and optimizing the parameters such as amplitude, frequency and phase of each of the constant number of waves is considered. Here we assumed the constant value to be $20$ for explanatory purpose. This summarizes the problem of optimizing the parameters frequency, amplitude and phase of each of the $20$ waves in that particular time frame of $1000$ samples using gradient descent algorithm assuming the step size as $1$ and considering squared error as loss function.

One can always experiment upon different optimizing algorithms and consider different values for the hyperparameters mentioned according to the audio data they have. We have also normalized the time frame values which act as input by dividing each value on time axis with $44100$ and then subtracting the mean from the input array and finally dividing it with the standard deviation. Two approaches have been followed to actually perform regression analysis as described above. The first approach is simple way of optimizing all the parameters of a particular time frame simultaneously at each step of gradient descent  \cite{ruder2016overview}. But, this method forces the waves to learn independently of each other which results in same optimized parameters for each wave. For example if frequency is $1$, amplitude is $1$ and phase is $0$ for the first wave in the hypothesis after optimizing, then, the each of the remaining $19$ waves of that time frame will also have the same values for frequency, amplitude and phase respectively. From first approach one can easily understand that the conventional form of regression analysis cannot be performed for music signal and hence, we have considered a second approach which is to optimize the second wave with respect to first, third with respect to second and first, and so on, similar to cost functions described by Algorithm \ref{algo1}. 
\begin{algorithm}[H]
	\caption{Optimization}\label{algo1}
		\begin{algorithmic}[1]
		 \State {\bfseries Input:} data $x$, size $n$; amplitudes $y$, size $n$; step $s$; starting index of time frame $start$
		\State {\bfseries Input:} parameters $param$,size $(20,3)$
		\State Initialize $h = array(zeros(1000))$.
		\For{$k=0$ {\bfseries to} $19$}
		\For{$j=0$ {\bfseries to} $9$}
		\For{$i=start$ {\bfseries to} $start+1000$}
		\State Assign \State$Ga=step*(h_i+param_{k_0}*\sin(2pix_i)-y_i)*(\sin(2\pi*x_i))$
		\State Assign
		\State $Gf = step*(h_i+\sin(2\pi*param_{k_1}*x_i)-y_i)*(2\pi*x_i\cos(2\pi*param_{k_1}*x_i))$
		\State Assign
		\State$Gp = step*(h_i+\sin(2\pi*x_i+2\pi*param_{k_2})-y_i)*(2\pi\cos(2\pi*x_i+2\pi*param_{k_2}))$
		\State Assign $param_{k_0} = param_{k_0}-Ga$
		\State Assign $param_{k_1} = param_{k_1}-Gf$
		\State Assign $param_{k_2} = param_{k_2}-Gp$
		\EndFor
		\EndFor
		\For{$v=start$ {\bfseries to} $start+999$}
		\State Assign $w = v \mod 1000$
		\State Assign
		\State $h_w = h_w + (param_{k_0} * np.sin(2 * np.pi * (param_{k_1} * x_v + param_{k_2})))$
		\EndFor
		\EndFor
	\end{algorithmic}
\end{algorithm}

As shown in the Algorithm \ref{algo1} we can update array $h$ which stores the superposition values of all $i$ number of waves while optimizing $i+1$ waves parameters. Thus, the superposition value can be added to redefine the cost function for each wave pertaining to the same time frame and thereby, optimizing the parameters of each wave with respect to the values obtained by the superposition of previous waves. 

\begin{figure}[htp]
	\begin{minipage}[b]{.5\textwidth}
		\centering
		\includegraphics[width=.85\textwidth]{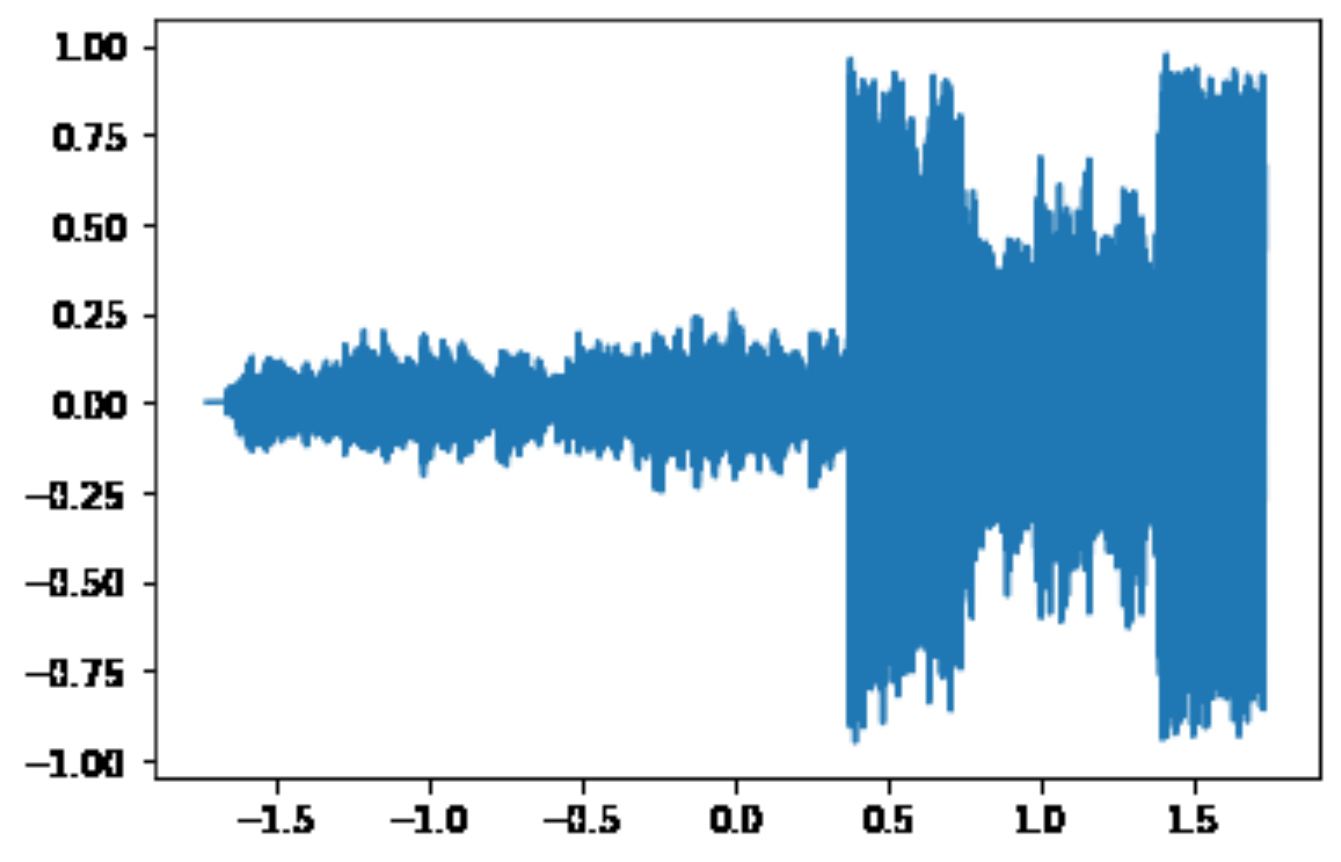}
		\caption{Original audio data with\\ desired amplitudes on y-axis vs. time \\period on x-axis}\label{9}
	\end{minipage}%
	\begin{minipage}[b]{.5\textwidth}
		\centering
		\includegraphics[width=.79\textwidth]{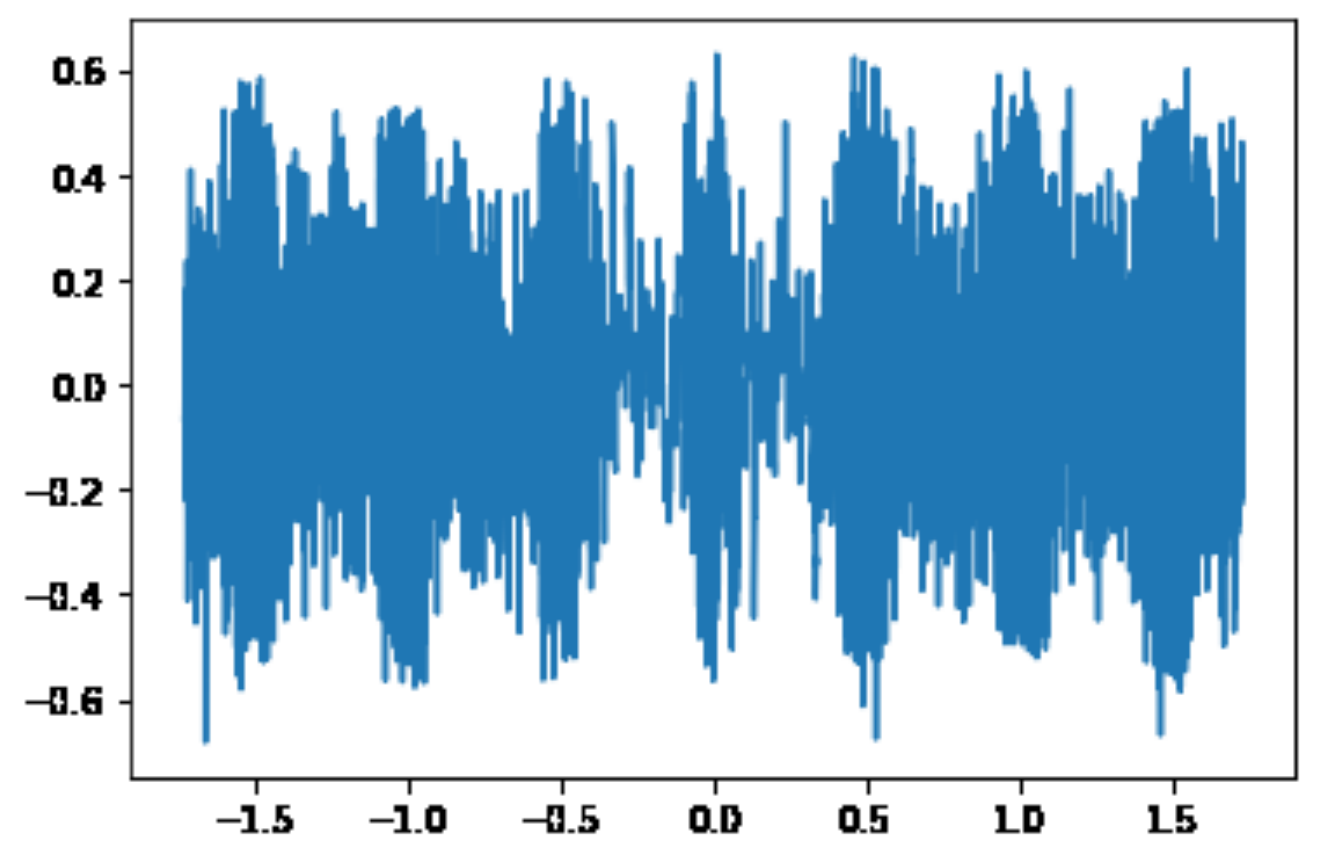}
		\caption{Plot with predicted amplitudes on y-axis vs. time period on x-axis by following independent parameter training excluding amplitude parameter.}
		\label{10}
	\end{minipage}
	\begin{minipage}[b]{.5\textwidth}
		\centering
		\includegraphics[width=.85\textwidth]{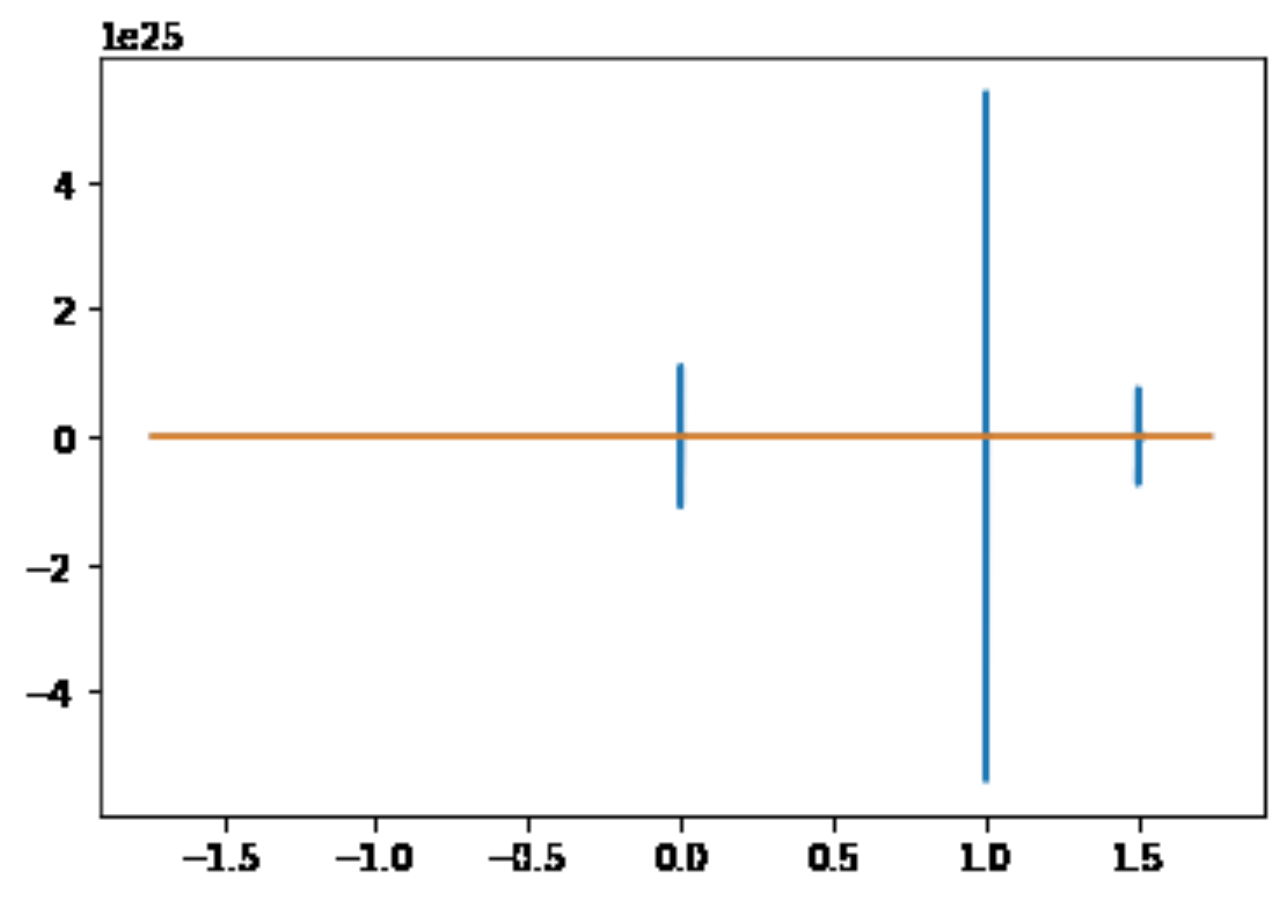}
		\caption{ Plot with predicted amplitudes on y-axis vs. time period on x-axis by following independent parameter training including \\amplitude parameter where horizontal plot \\ represents original signal.}
		\label{11}
	\end{minipage}
	\begin{minipage}[b]{.5\textwidth}
	\centering
	\includegraphics[width=.85\textwidth]{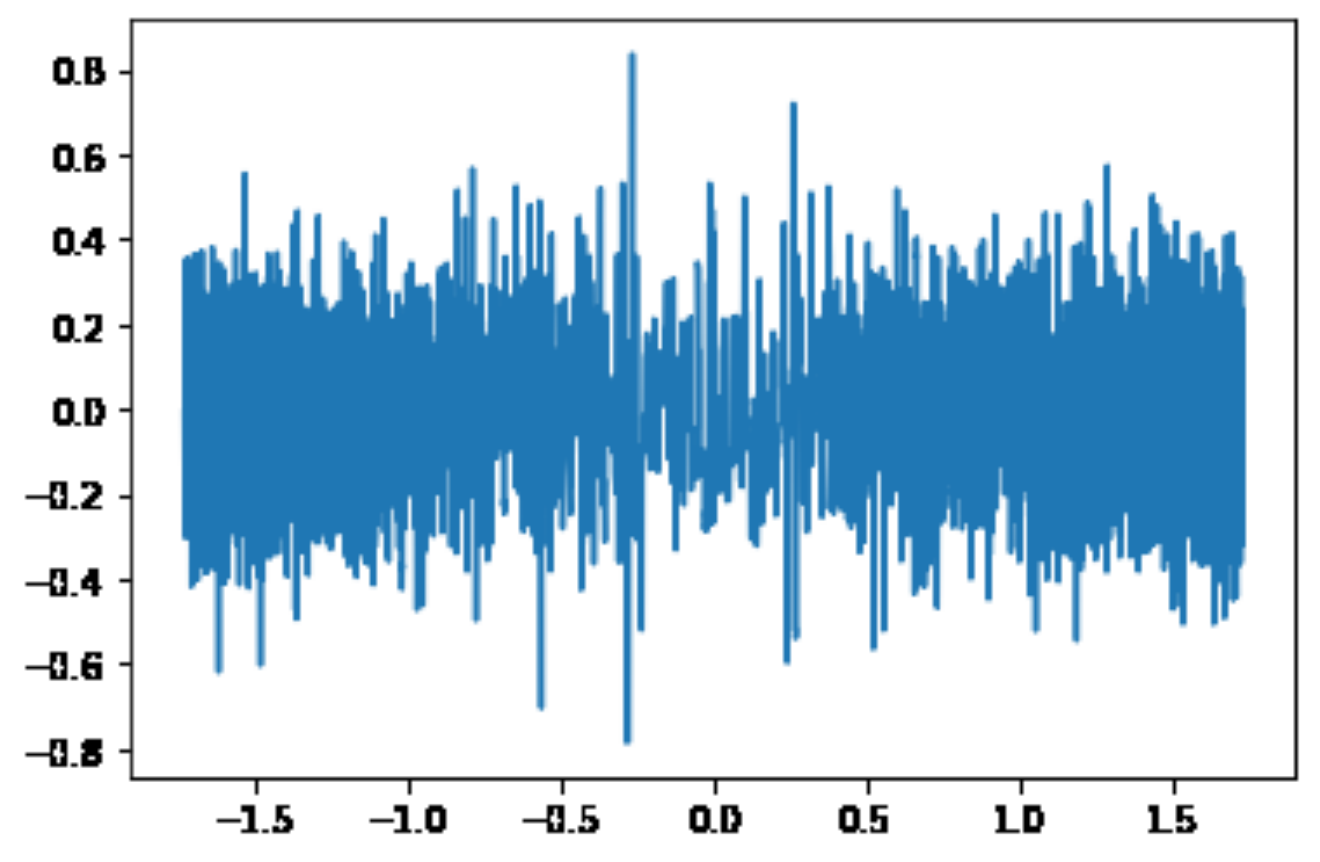}
	\caption{ Plot with predicted amplitudes on y-axis and time period on x-axis by following dependent parameter training.}
	\label{12}
\end{minipage}
\end{figure}

The fig.\ref{9} represents the graph of desired amplitudes vs. the time, and fig.\ref{10} shows the plot obtained by the hypothesis considered which is the superposition of 20 sine waves. Noted that the plot  in fig.\ref{10} is obtained by calculating amplitudes using the equation:
\begin{equation}\label{4}
	y = \sum_{i=0}^{20} a_i\sin(2\pi(f_ix + phase_i))
\end{equation}
by ignoring the amplitude parameter of each sine wave of that time frame as they were not even close to the desired values and scaling up the error by large extent which can be observed in fig.\ref{11}. This is a drawback with this approach which can be overcome by choosing a different optimization algorithm for amplitude parameter.

We considered the gradients for optimizing amplitude or frequency or phase  as follows:
\begin{align}
	&Ga = step * (h+a_i \sin(2\pi x)-y)*(\sin(2\pi x))\\
	&Gf = step * (h+\sin(2\pi f_i x)-y)*(2\pi x \cos(2\pi f_i x))\\
	&Gp = step * (h+\sin(2\pi x + 2 \pi p_i)-y)*(2\pi\cos(2\pi x+2\pi p_i))
\end{align}
where, $Ga$ is the amplitude gradient, $Gf$ is the frequency gradient, and $Gp$ is the phase gradient. Noted that, we only consider the effect of the parameters for which we compute the gradient. For example, while computing the gradient for amplitude parameter we consider $f_i$ as $1$ and $p_i$ as $0$ and thereby  optimizing only amplitude with respect to the samples, which is to try fit amplitude parameter for that wave for that time frame completely. Similar pattern can be observed for frequency where $a_i$ is made $1$ and $p_i$ as $0$ and in case of phase gradient $a_i$  and $f_i$ are both  $1$. This can be understood as independent parameter training for which we got the results as shown in fig.\ref{10}. 

We have also considered dependent parameter training where we try to optimize one with respect to other, for which the amplitude gradient $Ga$, frequency gradient $Gf$  and the phase gradient $Gp$ are:
\begin{align}\label{5}
	&Ga = step * (h+a_i \sin(2\pi f_i x)-y)*(\sin(2\pi f_i x))\\ 
	&Gf = step * (h+\sin(2\pi f_i x)-y)*((2\pi x)*\cos(2\pi f_i x))\\ 
	&Gp = step*(h+a_i\sin(2\pi f_i x+2\pi p_i)-y)*(2\pi a_i\cos(2\pi f_i x+2\pi p_i))
\end{align}
Here, the frequency is computed independently and amplitude is computed with respect to frequency parameter and finally phase parameter is computed with respect to both frequency and amplitude parameters. For dependent parameter training we observed a higher loss and hence, currently independent parameter training is better. Since we have not predicted the amplitude parameter for 20 waves of each time frame properly, we have divided the final value by 20 which is the mean amplitude at that particular instant. The figure for dependent parameter training can be seen in fig.\ref{12} and  observing fig.\ref{11}, one can calculate amplitudes by considering amplitude parameter for each of 20 waves in that time frame and clearly decide why we did not consider amplitude parameter.

\section{Conclusion}

Regression algorithm is the most fundamental and important algorithm which can be powerful when hypothesis, optimization and features are selected properly. It has the potential to even perform better than the current advanced machine learning techniques. With this theory we try to propose that, as algorithm selection is important for an application, similarly, data preprocessing and hypothesis reformulation is also that much important. We need to focus on formulating the underlying functions in preprocessing stage itself so that even on less amount of data, the algorithm can perform much more efficiently and we can eliminate the risks such as underfitting or overfitting. This also specifies that we need to conduct more  experiments with each algorithm by reformulating some of its parts on the dataset, so that, we can understand some of the relationships in the dataset and even have a combination of different machine learning algorithms acting on same dataset which may be much more efficient, and also understand the power of interdisciplinary algorithms. This also sheds light on the fact that adding features by exploring dataset can boost algorithm’s performance and efficiency.


\end{document}